\documentclass{emulateapj}

\shorttitle{CTCV\,J2056-3014: an unusual intermediate polar system}
\shortauthors{Lopes de Oliveira et al.}

\begin{document}

\title{CTCV\,J2056-3014: An X-ray-faint Intermediate Polar Harboring An Extremely Fast-spinning White Dwarf}

\author{R. Lopes de Oliveira}
\affiliation{Departamento de F\'isica, Universidade Federal de Sergipe, Av. Marechal Rondon, S/N, 49000-000 S\~ao Crist\'ov\~ao, SE, Brazil}
\affiliation{Observat\'orio Nacional, Rua Gal. Jos\'e Cristino 77, 20921-400, Rio de Janeiro, RJ, Brazil}

\author{A. Bruch}
\affiliation{Laborat\'orio Nacional de Astrof\'isica, Rua Estados Unidos, 154, CEP 37504-364, Itajub\'a, MG, Brazil}

\author{C. V. Rodrigues}
\affiliation{Divis\~ao de Astrof\'sica, Instituto Nacional de Pesquisas Espaciais, 12227-010, S\~ao Jos\'eS dos Campos, SP, Brazil}

\author{A. S. Oliveira}
\affiliation{IP\&D, Universidade do Vale do Para\'iba, 12244-000, S\~ao Jos\'e dos Campos, SP, Brazil}

\author{K. Mukai}
\affiliation{CRESST II and X-ray Astrophysics Laboratory, NASA Goddard Space Flight Center, Greenbelt, MD 20771, USA}
\affiliation{Center for Space Science and Technology, University of Maryland, Baltimore County, 1000 Hilltop Circle, Baltimore, MD 21250, USA}

\begin{abstract}


We report on XMM-{\it Newton} X-ray observations that reveal CTCV J2056-3014 to be an unusual accretion-powered, intermediate polar (IP) system. 
It is a member of the class of X-ray-faint IPs whose space density remains unconstrained but potentially very high, with L$_{x,0.3-12 keV}$ of 1.8$\times$10$^{31}$\,erg\,s$^{-1}$. 
We discovered a coherent 29.6\,s pulsation in X-rays that was also revealed in our reanalysis of published optical data, showing that the system harbors the fastest-spinning, securely known white dwarf (WD) so far. There is no substantial X-ray absorption in the system.  Accretion occurs at a modest rate ($\sim$\,6$\times$10$^{-12}$\,M$_{\sun}$\,yr$^{-1}$) in a tall shock above the WD, while the star seems to be spinning in equilibrium and to have low magnetic fields. Further studies of CTCV J2056-3014 potentially have broad implications on the origin of magnetic fields in WDs, on the population and evolution of magnetic cataclysmic variables, and also on the physics of matter around rapidly rotating magnetic WDs.  

\end{abstract}


\section{Introduction} \label{sec:intro}

\object{CTCV J2056-3014} (hereafter J2056) is a cataclysmic variable (CV) with an orbital period of 1.76\,hr determined from time-resolved optical spectroscopy \citep{2010MNRAS.405..621A}. 
Based on their detection of a short periodicity at 15.4 min from optical photometry and the fact that the system matches a relatively bright X-ray emitter in the ROSAT (PSPC) Bright Source Catalog \citep[0.10$\pm$0.02\,cts\,s$^{-1}$ at 0.1-2.4\,keV;][]
{2009ApJS..184..138H}, \citet{2010MNRAS.405..621A} suggested J2056
to be an intermediate polar (IP) candidate, i.e., an asynchronously rotating magnetic white dwarf (WD) accreting matter from a Roche-lobe-filling donor usually via a partial accretion disk \citep{1994PASP..106..209P}.

\citet{2017AJ....153..144O} presented additional optical spectroscopy of the system and pointed out the similarity of its spectrum to the spectral features  found in  the  rare IPs with short orbital periods. These features include  H$\beta$ as intense as H$\alpha$ and weak He{\sc II} 4686~\r{A} in emission. 
Optical photometry of J2056 conducted by \citet{2018NewA...58...53B} revealed that the averaged magnitude of the system varies by at least 2.4-mag in comparison with measurements of \citet{2010MNRAS.405..621A}, ranging from V\,$\sim$\,17.6 to 15.2 mag, and that it displays strong flickering with an amplitude up to 0.8 mag.  \citet{2018NewA...58...53B} also suggested that the 15.4-min period claimed by \citet{2010MNRAS.405..621A} is  spurious. 
Finally, Gaia parallax indicates that J2056 is a nearby system, at a distance ($d$) of 261.6$\pm$7.4\,pc \citep{2018AJ....156...58B}.

We have started an XMM-{\it Newton} X-ray follow-up program for validation of CV candidates originally identified in optical surveys, which includes J2056. Here we report on its X-ray properties, which suggest J2056 to be an unusual IP. We also revisit the optical observations of \citet{2018NewA...58...53B}.

\section{Observations} \label{sec:style}

\subsection{X-ray data}

J2056 was observed for about 18\,ks on 2019 October 24 by  XMM-{\it Newton} (ObsID 0842570101; PI: R. Lopes de Oliveira). The snapshot was focused on X-ray spectrophotometry with the European Photon Imaging Camera (EPIC), namely MOS1, MOS2, and pn cameras. The Reflection Grating Spectrometers RGS1 and RGS2 did not collect enough photons to allow us to carry out high-resolution X-ray spectroscopy.
The UV observations with the optical monitor in timing mode barely covered the source position, rendering its data unusable for timing analysis. 

The EPIC observations were partially contaminated by solar particles, with only a low background level during about 15.2\,ks for the MOS cameras, and 8.7\,ks for the pn camera. No pile-up or technical issues were identified in these data. 
The observations were reduced and data products were extracted following standard procedures using the Science Analysis System (SAS) v18.0.0. In particular, they were reprocessed using the \textsc{epproc} (for the pn data) and \textsc{emproc} (for the MOS1-2 data) tasks.
We used calibration files downloaded on 2020 January 2. 
Spectral analysis was accomplished using the \textsc{xspec} software version 12.9.1m.

\subsection{Optical data}
\label{sct:opticaldata}

We revisited the optical photometric observations of J2056 presented by \citet{2018NewA...58...53B} with the sole purpose of checking for high-frequency periodicities. This effort was motivated by the detection of pulsation in X-rays (Section \ref{sct:photometry}). The optical observations were carried out on four nights in 2015 (June 9-12) and on two nights in 2016 (September 7-8) with the 0.6m Zeiss telescope of Observat\'orio do Pico dos Dias -- Laborat\'orio Nacional de Astrof\'{\i}sica, Brazil.
Light curves spanning from 55 to 340 minutes were obtained with a time resolution of 5\,s.  To maximize the count rates within the short exposures, no filter was used. The throughput of the instrumentation corresponded roughly to V magnitude \citep{2018NewA...58...53B}. 
Basic data reduction (bias removal, flat-fielding) and aperture photometry were performed using default procedures with IRAF \citep{1986SPIE..627..733T} and with the MIRA (Bruch, 1993, ``MIRA: A Reference Guide'', Univ. Münster) software system, respectively.

\begin{table*}
\begin{center}
\scriptsize
\caption{The best-fit spectral X-ray parameters} \label{tbl:spct_xrays}
\begin{tabular}{lcccccccc}
\tableline\tableline
                            & N$_{H}$ & $k$T$_{\rm apec}$              & $k$T$_{\rm apec}$             & $k$T$_{max,{\rm mkcflow}}$       & $Z$  & $\chi^{2}_{\nu}$/d.o.f. &  unabs. f$_{(0.3-12\,keV)}$ & L$_{0.3-12\,keV}$\\
                            & (10$^{20}$\,cm$^{-2}$) & (keV)             & (keV)             &  (keV)      & ($\times\,Z_{\odot}$) & &  (erg\,cm$^{-2}$\,s$^{-1}$) & (erg\,s$^{-1}$) \\
\tableline
\textsc{phabs$*$(apec+apec)}    & 2.4$^{+0.6}_{-0.6}$            & 0.81$^{+0.02}_{-0.02}$ & 5.11$^{+0.24}_{-0.24}$ & ...                     & 0.74$^{+0.14}_{-0.13}$  & 1.19/386    & 2.1$\times$10$^{-12}$  & 1.7$\times$10$^{31}$($d$/261.6\,pc)$
^{2}$          \\
\textsc{phabs$*$mkcflow}      & 2.9$^{+0.5}_{-0.5}$            & ...                    & ...                    & 9.88$^{+0.35}_{-0.39}$  & 0.80$^{+0.10}_{-0.09}$  & 1.18/388 & 2.1$\times$10$^{-12}$ & 1.7$\times$10$^{31}$($d$/261.6\,pc)$
^{2}$               \\
\textsc{phabs$*$(apec+mkcflow)} & 2.7$^{+0.7}_{-0.7}$            & 0.79$^{+0.04}_{-0.04}$ & ...                    & 14.34$^{+1.18}_{-1.21}$ & 0.81$^{+0.16}_{-0.14}$  & 1.09/386  & 2.2$\times$10$^{-12}$ & 1.8$\times$10$^{31}$($d$/261.6\,pc)$
^{2}$             \\
\tableline
\end{tabular}
\end{center}
\end{table*}

\section{Results}

\subsection{X-ray spectroscopy}
\label{X-ray spectroscopy}

The net count rates of J2056 at 0.3--12\,keV were 0.194$\pm$0.004 counts\,s$^{-1}$, 0.202$\pm$0.004 counts\,s$^{-1}$, and 0.739$\pm$0.010 counts\,s$^{-1}$ for the MOS1, MOS2, and pn cameras, respectively. The spectra were binned such that each bin had at least 25 counts and thus the $\chi^2$ method was applied to both fit and test statistics in their modeling with \textsc{xspec}.

\begin{figure}
\centerline{
\includegraphics[angle=-90,scale=0.32]{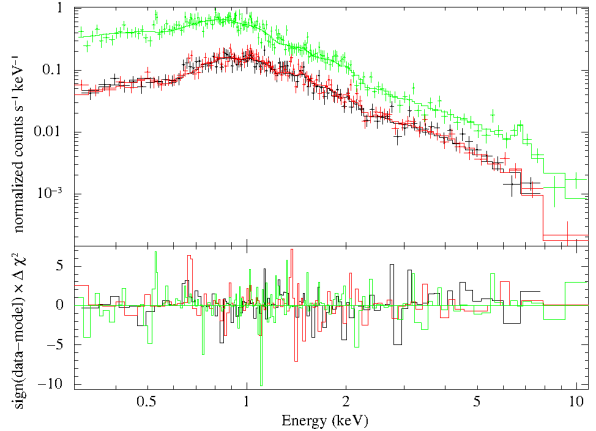}
}
\caption{X-ray spectra (top) and residuals (bottom): black, red, and green colors correspond to MOS1, MOS2, and pn data, respectively. The continuous lines are the single thermal plus cooling-flow emission model fits to the data.} \label{fig:spct}
\end{figure}

Even with relatively short exposures, 
the observations resulted in good-quality EPIC spectra (Fig. \ref{fig:spct}). The X-ray energy distribution of J2056 extends over the whole energy range covered by the EPIC cameras. An excess emission due to ionized lines of the Fe\,K$\alpha$ complex at 6.6--7\,keV is seen in the pn data and supports the interpretation, as usual in accreting WDs, of the predominantly thermal nature for the X-ray emission. There is no evidence of an optically thick, blackbody-like component from the WD surface. Thus, we applied the following \textsc{xspec} models that account for the emission from collisionally ionized diffuse gas due to accretion as the primary energy source: \textsc{apec}, describing a single thermal plasma, and \textsc{mkcflow}, a multi-temperature model representing a cooling-flow \citep{1988ASIC..229...53M}. The \textsc{mkcflow} model was interpolated by using the AtomDB \citep{2012ApJ...756..128F} data so that it is equivalent in assumptions to those used in the \textsc{apec} model. We adopted the abundance table of \citet{2009ARA&A..47..481A}. The \textsc{phabs} model was used to account for the photoelectric absorption effects on X-rays.

A single temperature component (\textsc{phabs$*$apec}) does not match the spectra well, failing to explain the continuum below 2\,keV and the Fe\,K$\alpha$ complex ($\chi^2_{\nu}$\,=\,2.18). The inclusion of a second thermal component (\textsc{phabs$*$(apec+apec)}) improves the fit but fails especially in the description of the 0.9-1.5\,keV region and slightly overpredicts the Fe\,K$\alpha$ lines ($\chi^2_{\nu}$\,=\,1.19). Despite similar statistics ($\chi^2_{\nu}$\,=\,1.18), the cooling flow model
(\textsc{phabs$*$mkcflow}) visually improves the description of both continuum and iron lines. Finally, an acceptable description ($\chi^2_{\nu}$\,=\,1.09) is found by adding a single thermal component to the cooling flow model, providing a good fit of the 1\,keV region, which is expected to be rich in unresolved emission lines. 
As for the \textsc{mkcflow}, we fixed the unconstrained $k$T$_{low}$ parameter to its minimum value of 0.0808\,keV and the required redshift parameter to 6.1$\times$10$^{-8}$ from the Gaia distance and standard cosmological values of \textsc{xspec}.
Table \ref{tbl:spct_xrays} lists the best-fit spectral parameters of the models described above.
Figure \ref{fig:spct} shows the EPIC spectra and the final model \textsc{phabs$*$(apec+mkcflow)}; henceforth, this is the model discussed in this letter. 

Our best-fit model indicates that the X-rays are absorbed by the equivalent in hydrogen column density (N$_{H}$) of 2.7$^{+0.7}_{-0.7}\times$10$^{20}$\,cm$^{-2}$. The X-ray emission is dominated by a moderately hard thermal component that cools down from $k$T\,=\,14.34$^{+1.18}_{-1.21}$\,keV. A secondary contribution is well described by a plasma component having $k$T\,=\,0.79$^{+0.04}_{-0.04}$\,keV, which accounts for about 6.2\% of the total unabsorbed flux at 0.3-12\,keV. A subsolar abundance of 0.81$^{+0.16}_{-0.14}$\,$Z_{\odot}$ is inferred, forced to be the same for both thermal components during the fits, but its determination strongly depends on the Fe\,K$\alpha$ lines. From the \textsc{mkcflow} component, the mass accretion rate is  5.9$^{+0.5}_{-0.4}\times$10$^{-12}$\,M$_{\sun}$\,yr$^{-1}$. The total luminosity of the system at 0.3-12\,keV is 1.8$\times$10$^{31}$($d$/261.6\,pc)$
^{2}$\,erg\,s$^{-1}$. These results are discussed in Section \ref{sct:discussion}.

\subsection{X-ray and optical timing analysis}
\label{sct:photometry}

Time flags were converted to the Barycentric Dynamical Time scale using the online tool\footnote{See http://astroutils.astronomy.ohio‑state.edu/time} of \citet{2010PASP..122..935E} for the optical observations and the \textsc{barycen/SAS} task for the X-ray data. 
Background-corrected X-ray light curves from each EPIC camera were produced considering a binning of 10\,s. 
We considered three energy ranges:
 0.3--10\,keV, 0.3--2\,keV (``soft''), and 2--10\,keV (``hard''), in order to access the energy dependence of any variable signal. 
The light curves of each EPIC camera were investigated separately. Optical light curves were constructed retaining the original resolution of 5\,s.

The search for periodicities was carried out using the 
Lomb-Scargle periodogram \citep{1976Ap&SS..39..447L,1982ApJ...263..835S} in optical and X-ray light curves. 
We explored X-ray light curves considering two datasets. The first one incorporates the entire observations. It includes spikes in background count rates, which can be as high as the source signal, especially in the 2--10\,keV band. This condition lasted for about 2.6\,ks. The second one considers only data that were collected during the last $\sim$ 7.5\,ks, the longest continuous time interval with low particle background.

\begin{figure}
\centerline{
\includegraphics[angle=-0,scale=0.38]{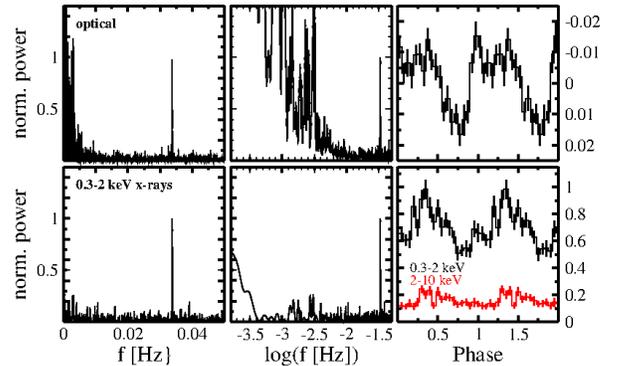}
}
\caption{Periodograms and folded light curves on the 29.6\,s period from the optical (top) and X-ray (bottom; from pn) data. For X-rays, the periodograms corresponding to the soft band and light curves are shown for the soft and hard bands. 
} \label{fig:lc}
\end{figure}

A high-significance peak associated with a period of 29.6\,s is clearly seen in the Lomb-Scargle periodogram applied to the 0.3-10\,keV and 0.3-2\,keV light curves of all EPIC cameras (Fig. \ref{fig:lc}). As for the 2-10\,keV band, the peak is recovered only in light curves produced considering a time interval with low background contamination, and with a lower power when compared to results of the other two energy ranges. This is mainly due to the low signal-to-noise ratio in the hard energy range. 

As reported by \citet{2018NewA...58...53B}, the optical light curve of J2056 displays variations on timescales of hours superposed by flickering.
A search for high-frequency periodicities in optical data was first carried out individually for each of the six nights. 
The 29.6\,s pulsation seen in X-rays is also identified in all optical light curves, and the periods in both regions are the same at the level of the formal errors. The upper left and center panels of Fig. \ref{fig:lc} show the periodogram of the 2015 June 10 optical light curve. The optical data taken in subsequent nights do not indicate migration of phase. 
We refine the periodicity in the optical by merging the light curves of each of the two observing seasons, resulting in an average period of 29.6098$\pm$0.0014\,s.\footnote{Note that the time difference between the two seasons is too large to concatenate the two data sets without cycle count ambiguities and thus to refine the period even more.} The same value was derived from the 0.3-2 keV pn light curve considering the entire observation, but with an uncertainty of $\pm$0.0213\,s. 
 
Figure \ref{fig:lc} (right frames) shows folded optical and X-ray light curves. 
Before folding the optical data on the 29.6 s period, variations on longer timescales 
were removed from the light curve by subtraction of a filtered version generated by the Savitzky-Golay algorithm \citep{1964AnaCh..36.1627S} which eliminates variations below a cutoff timescale, here chosen to be 1 min.
We adopted a conservative approach and used the free-flare dataset to construct the phase-folded, X-ray light curves. It still covers $\sim$250 pulsation cycles and avoids features that may be background-induced.

The waveform in the optical phase diagram is characterized by a deep minimum and two maxima of approximately equal height, separated by about 0.4 in phase (see the upper right frame of Fig. \ref{fig:lc}). The semi-amplitude of these variations is $\sim$\,0.016 mag, and its shape is quite similar to that observed in soft X-rays (0.3--2\,keV), except for the difference in height of the X-ray maxima. The pulsed fraction is significant in all light curves. 
In the soft band, the variability reaches about 25\% of the mean level.
A similar value is obtained for the integrated band because the total counts are dominated ($\sim$80\%) by soft X-ray photons. 
As for ``hard'' (2-10\,keV) X-rays, even though the pulsation is not as clearly observed in the periodograms, the folded light curve indicates that the 29.6\,s modulation is present and has a significant ($\sim$\,50\%) pulsed fraction (see Fig. \ref{fig:lc}).

Optical and X-ray modulations can be understood within the same scenario. The stability of the 29.6\,s modulation over two years in the optical (2015 and 2016) and its presence three years later in X-rays (2019) lead us to the interpretation that it represents the spin period of the WD in J2056. 

\section{Discussions and conclusions}
\label{sct:discussion}

Our main findings are: (i) J2056 is an IP harboring a fast-spinning WD, and (ii) its X-ray luminosity is low for an IP. Those properties together with the short orbital period \citep[1.76\,hr;][]{2010MNRAS.405..621A}, below the CV orbital period gap and rare among IPs, make J2056 an unusual and interesting IP.

Many IPs have been discovered through INTEGRAL and Swift/BAT hard X-ray (E\,$>$\,10 keV) surveys \citep{2019arXiv190906306D}, for which interstellar absorption is not an issue. 
They have typically L$_{\rm X}$ above 10$^{33}$\,erg\,s$^{-1}$ 
meaning that, with the available sensitivities, the systems can be discovered out beyond 1\,kpc \citep{2014MNRAS.442.2580P}. Although large enough to yield a statistically significant sample of luminous IPs, the hard X-ray source catalogs are small enough ($<$\,2000 objects) for systematic identification and follow-up programs \citep[see, e.g.,][]{2018AJ....155..247H,2019arXiv190906306D}.
However, there appears to be a separate class of low-luminosity IPs \citep[LLIPs;][]{2014MNRAS.442.2580P} typically with L$_{\rm X}$\,$\sim$10$^{31}$\,erg\,s$^{-1}$, only a subset of which have been detected in hard X-ray all-sky surveys \citep{2017PASP..129f2001M}. The LLIP population seems to be dominated by short orbital period systems \citep{2014MNRAS.442.2580P}, but there is no known unique set of characteristics that allows us to readily identify its members to construct a distance-limited, complete sample.
As argued by \citet{2014MNRAS.442.2580P}, the separate and yet not constrained LLIP population may be numerous enough to match the common IPs in integrated X-ray luminosity. Thus, many LLIPs may be awaiting discovery. In this context, the identification of J2056 as an LLIP is significant.

J2056 has the fastest-spinning WD among confirmed IPs and also holds the record of all securely known WDs. We are aware of only two systems that may be harboring a WD rotating faster than J2056: WZ\,Sge and RX\,J0648.0-4418. WZ\,Sge exhibits intermittent modulations at 27.87\,s and 28.96\,s, which may be associated with the rotation of its WD, but this identification is not secure \citep[see results and discussion in][] {2014A&A...566A.121N}. 
RX\,J0648.0-4418 contains a 1.28\,M$_\odot$ \citep{2009Sci...325.1222M} compact object spinning at 13.2\,s \citep{1997ApJ...474L..53I} but its nature is not clear: it may be a WD in an early evolutionary stage \citep{2018MNRAS.474.2750P} or a neutron star. \citet{2016MNRAS.458.3523M} argued that the spin-up rate derived for RX\,J0648.0-4418 (2.15$\times$10$^{-15}$\,s\,s$^{-1}$) would be unusual for an accreting WD and this scenario would be strongly disfavoured if the distance is confirmed to be less than $\sim$\,4\,kpc. And, in fact, the system seems to be much nearer: $d_{Gaia}$\,=\,501.1$_{-15.6}^{+16.7}$\,pc \citep{2018AJ....156...58B}. Moreover, we notice that the bulk of the X-ray emission is interpreted as being due to a non-thermal (power-law) component \citep{2013A&A...553A..46M}, which is not expected for an accreting WD but it is the rule for accreting neutron stars

Two lines of argument strongly suggest that the magnetic field of
J2056 is low for an IP. The first is to assume that J2056, as is
likely for IPs as a group \citep{Patterson_2020}, is in spin equilibrium. Consideration of
material torques (which act to spin up the WD) and magnetic
torques (which must balance the material torques in equilibrium) leads
to the conclusion that the magnetic fields of fast-spinning IPs, such
as J2056, are lower than their longer spin period cousins (see
equation 21 and Figure 17 of \citealt{1994PASP..106..209P}). 
This assumption leads to a preliminary
estimate of magnetic moment $\mu\sim$ 5$\times10^{30}$\,G\,cm$^3$ for J2056 but this needs to be
revisited after further studies. Note that this is not a conclusion that applies to all LLIPs, 
because at least two systems have relatively long spin periods -- 
EX\,Hya with 4021.6\,s \citep[e.g.,][]{2009IBVS.5876....1M} and V1025\,Cen with 2146.59\,s \citep{1998MNRAS.299...83B}.
The second argument, instead of the spin equilibrium assumption, relies on
the fact that accretion is suppressed when the inner edge of the disk
is rotating more slowly than the magnetic field lines. Such systems
are believed to behave as magnetic propellers 
 \citep[see the case of AE\,Aqr:][]{1997MNRAS.286..436W}. Since J2056 is entirely consistent with being
accretion-powered and does not display any signatures of a propeller,
the Keplerian frequency at the inner edge of its disk must be
$\sim$29.6 s or shorter. Such a small magnetospheric radius, combined
with the modest accretion rate (Section \ref{X-ray spectroscopy}), demands a low magnetic field.

The X-ray emission can be explained by the cooling flow framework expected for accretion-powered WDs, plus a single thermal plasma that responds for about 6\% of the total luminosity (Section \ref{X-ray spectroscopy}). 
The latter contribution likely represents a marked deviation from the assumptions behind the \textsc{mkcflow} model, which are expected to be violated in the condition of a tall shock along which the accreting matter is expected to suffer non-negligible gravitational acceleration. In fact, the luminosity and therefore the accretion rate per unit area is low, suggesting that the shock is not occurring near the WD surface. Under these conditions, the often-made assumptions of radial accretion, freefall from infinity, and shock near the WD surface are not applicable and thus the maximum temperature of the shock  (in this case $k$T\,=\,14.34$^{+1.18}_{-1.21}$\,keV) cannot be directly used to determine the mass of the WD (which under such assumptions would be around 0.46\,M$_{\odot}$).

The tridimensional extinction map of 
\citet{2019A&A...625A.135L} 
combined with a Gaia distance of 260\,pc suggest a best-guess 
E(B-V) of 0.014 mag that corresponds to 
N$_{H}$\,$\sim$\,1.2$\times$10$^{20}$\,cm$^{-2}$, 
but one as high as 0.033 mag indicating 
 N$_{H}$\,$\sim$\,2.5$\times$10$^{20}$\,cm$^{-2}$ in the line of sight to 
J2056 is still possible. The value inferred from X-rays (Section \ref{X-ray spectroscopy}) is
N$_{\rm H}$\,=\,2.9$^{+0.5}_{-0.5}\times$10$^{20}$\,cm$^{-2}$, which is at most only slightly higher than that due to the interstellar medium. 
Thus, contrary to what is typical in luminous IPs, there is no significant intrinsic X-ray absorption in J2056. 

The lack of a strong intrinsic (complex) absorber for J2056 
is another common characteristic of LLIPs \citep[see, for example, the case of DW Cnc:][]{2019MNRAS.484.3119N}. The low absorption in J2056 is an additional piece of evidence for a tall shock. This is because tall shocks allow us to see the X-rays from the side of the post-shock region with no expectation of being affected by a complex absorber. The opposite is a common characteristic of classic (luminous) IPs, in which the shock is near the WD surface and our lines of sight almost inevitably cross a strong complex absorber in the pre-shock flow. 

J2056 is entirely accretion-powered, which is an important finding if compared with other objects with similar spin periods. AE Aqr, a peculiar IP harboring a WD with a spin period of 33\,s (and orbital period of 9.88\,hr) and with an even lower X-ray luminosity than J2056, is thought to be in the propeller regime \citep{1998MNRAS.298..285W}.
AR Sco has a WD with a spin period of 117\,s (orbital period of 3.56\,hr) and may even be entirely rotation-powered, as a ``white-dwarf pulsar" \citep{2016Natur.537..374M,2017NatAs...1E..29B}.

The example of J2056, displaying X-rays that are not luminous or hard enough to have attracted attention in previous surveys, leads to a promising strategy to identify further LLIPs by follow-up X-ray observations of short orbital period CVs and candidates. 
The eROSITA survey will likely reveal the true extent of the LLIP
population by measuring the X-ray fluxes of all known CVs and
discovering many new ones \citep{2019cwdb.confE..49S}.
This is important for advancing our understanding of the physics and the evolutionary history of these systems. In fact, J2056, with its  low X-ray luminosity and the fast spin of its WD, may be typical of a currently unrecognized sub-population of the class. If that is the case, implying that there is a large population of lower magnetic field IPs, this is an important clue that must be factored into the theory of the origin of the magnetic field in WDs, and that of the population and evolution of magnetic CVs. Moreover, J2056 offers an important test case for the physics of matter around a rapidly rotating magnetic WD. 

\acknowledgments

C.V.R. thanks the grants \#2013/26258-4, S\~ao Paulo Research Foundation (FAPESP), and \#303444/2018-5, CNPq. A.S.O. acknowledges São Paulo Research Foundation (FAPESP) for financial support under grant \#2017/20309-7.

\vspace{5mm}

{\it Facilities:} \facility{XMM-Newton}, \facility{LNA:0.6-m Zeiss}.

\end{document}